\documentclass[aps,prl,preprint,groupedaddress]{revtex4-2}
\usepackage{graphicx}
\begin{document}

%\preprint{}

\title{Statistical hydraulic model for the Leonardo's rule}

\author{O. Sotolongo-Costa}
\email{osotolongo@uaem.mx}
\affiliation{%
Centro de Investigaci\'on en Ciencias, Instituto de Investigaciones en Ciencias B\'asicas y Aplicadas, Universidad Aut\'onoma del Estado de Morelos, Av. Universidad 1001, Col. Chamilpa, 62209 Cuernavaca, Morelos, Mexico.
}%

\author{P. Villasana-Mercado}
\affiliation{%
Unidad Acad\'emica de F\'isica, Universidad Aut\'onoma de Zacatecas, Calzada Solidaridad Esquina con Paseo la Bufa S/N, 98060 Zacatecas, Zac., Mexico. 
}%

\author{L. S\'anchez-Calder\'on}
\affiliation{%
Unidad Acad\'emica de Ciencias Biol\'ogicas, Universidad Aut\'onoma de Zacatecas, Calzada Solidaridad Esquina con Paseo la Bufa S/N, 98060 Zacatecas, Zac., Mexico. 
}%

\author{I. Rodr\'iguez-Vargas}%
 \email{isaac@uaz.edu.mx}
\affiliation{%
Unidad Acad\'emica de Ciencia y Tecnolog\'ia de la Luz y la Materia, Universidad Aut\'onoma de Zacatecas, Carretera Zacatecas-Guadalajara Km. 6, Ejido La Escondida, 98160, Zacatecas, Zac., Mexico
}%

\date{\today}

\begin{abstract}
\noindent More than five hundred years ago Leonardo Da Vinci found a pattern in the growth of trees nowadays known as the Leonardo's rule. This rule relates the thickness of the stem with the thickness of the branches at different bifurcation stages in a Pythagorean fashion. He argued that his rule was the result of the conservation of sap flux. In the present work, we explore this idea by assuming that the sap flux through each xylem element behaves as a non-ideal fluid and the size-distribution of the xylem elements obeys a power law distribution. We find that the simultaneous fulfillment of  Leonardo's rule and the conservation of the sap flux, lead to a global behavior of the sap like that of an ideal fluid after summing over all xylem elements. These results are supported by field and experimental work. In particular, we corroborated the Leonardo's rule in different tree species by measuring the stem-branches thickness at different bifurcations stages. We also determined the statistical size-distribution of the xylem elements through a maceration process, finding that it corresponds to a power law distribution with an exponent close to three, which is the exponent required for the Leonardo's rule. As far as we know this is the first time that a statistical hydraulic model supported by experimental data is presented for the Leonardo's rule.
\end{abstract}


\maketitle

\section{Introduction}

Leonardo da Vinci was a man with innate curiosity and exceptional intellectual capacity that allowed him to understand nature \cite{Capra2013}. Leonardo is best known for his iconic paintings the Mona Lisa and the Last Supper \cite{pevsner2002} as well as his remarkable drawing the Vitruvian Man \cite{suh2005}. However, his legacy goes beyond arts with seminal contributions to science. Based on observation and experience he deciphered different patterns in nature. In plants he unveiled the so-called sixth leaf, rings of growth and branching patterns in trees nowadays known as Leonardo's rule \cite{Capra2013,suh2005}. Regarding the latter, he found that the branches at different heights are proportional to the stem. In specific, the sum of the transverse areas of the branches at different bifurcation levels remains constant and equal to the transverse area of the stem. Leonardo, based on his knowledge of the transportation of water in plants, attributed this relationship to the conservation of sap flux \cite{Capra2013}. The modern version of the Leonardo's rule is given in Pythagorean fashion $d^{\Delta}=\sum^n_{i=1} d_i^{\Delta}$, where $d$ is the diameter of the stem, $d_i$ is the diameter of the $i$-th branch at a particular bifurcation level, $n$ the number of branches at the mentioned level and $\Delta$ the so-called Leonardo's exponent \cite{mandelbrot1982}. The nominal value of the Leonardo's exponent is 2, however it is well known that it depends on the tree species \cite{mandelbrot1982,aratsu1998,sone2008}. The Leonardo's rules is also obeyed by the radical system of trees \cite{armin2000}. Furthermore, the Leonardo's rule in conjunction with the external fractal architecture of trees is involved in the optimization of different biological processes \cite{eloy2011,eloy2017}. For instance, the modeling of trees shows that the mechanical stress induced by wind loads is minimized by the self-similar organization of the branches as well as the area-preserving condition \cite{eloy2011}. Likewise, the competition for light is optimized due to the complex-fractal growth of trees in which the Leonardo's rule is implicated \cite{eloy2017}. 

There are also several works assessing the impact of hydraulics in plant vascular systems \cite{shinozaki1964,aratsu1998,lehnbach2018,west1999,savage2010}. From the rather simple pipe model to the most sophisticated proposals based on internal and external fractal architectures. The common factor in all these studies is the derivation of well-known allometric scaling laws. In the case of the pipe model \cite{shinozaki1964,aratsu1998,lehnbach2018}, trees are modeled as a uniform tube from the base to the top, with the tube diameter equivalent to the cross-section of the stem. The pipe model entails the Leonardo's rule as well as the non-linear scaling relationships for the height and mass of a plant with respect to the basal stem diameter. The pipe model has also been useful to determine the biomass of plants. A historic review of the pipe model can be consulted in \cite{lehnbach2018}. The WBE model is a more refined proposal in which a general theory of resource distribution through hierarchical branching networks is considered \cite{west1999}. In particular, the model takes into account external space filling (self-similar branching), hydraulic optimization (tube tapering) and scale free canopy (invariant leaf-petiole size). In this model despite tubes taper the flow rate is size independent due to the increase of the flow velocity in small radius branches. The WBE model also shows that the well-known allometric scaling laws for animals \cite{schmidt1984} apply for plants \cite{niklas1994}, supporting the idea that they are universal in biology and can be modeled by simple geometric and hydrodynamic principles \cite{west1997}. A decade later of the WBE model a theory that also considers internal space filling as well as trade-offs between hydraulic safety and efficiency was proposed \cite{savage2010}. The interrelationship between the vascular and branching networks give rise to more general allometric scaling laws that can respond to the particularities of tree species. In addition, the model gives predictions for the sap (solutes and water) flow, the taper and frequency of xylem conduits. In contrast to the WBE model the flow velocity remains constant regardless of the tapering of the xylem conduits, and the flow rate size independent as a result of the internal fractal architecture. The area preserving condition is also fulfilled in this model. As we have documented, it is well accepted that the external branching network minimizes the mechanical stress in trees, optimizes the capture of light for the photosynthesis process and shapes the fundamental allometric scaling laws. However, for the internal vascular network there is no consensus and depending on the model we can have different outcomes, as in the case of the fluid velocity. Finally, it is important to remark that in all these models the area preserving condition (Leonardo's rule) and the size-independent flow rate (flow conservation) are obeyed. So, it is interesting to see how these conditions as fundamental determine by and large the size distribution of xylem conduits. 	

In this work we explore the seminal idea of Leonardo that the conservation of the sap flux is directly implicated in the Leonardo's rule. We propose a statistical hydraulic model based on a power law distribution for the xylem elements. We corroborate the Leonardo's rule by measuring the radius of the stem and branches of different tree species. We also obtain the size distribution of the xylem elements after subjected one of the species to a maceration process. A fairly good agreement is obtained between the theoretical prediction and the experimental results for the size distribution of the xylem elements. 

\section{Statistical hydraulic model}
As we have documented the hydraulic model has been modified and refined throughout years. From the rather simple pipe model to the more elaborated proposals of the last decades. In the present work, we adopt an statistical hydraulic version based on a power law distribution for the xylem elements that accounts of the space filling internal fractal architecture of trees as well as memory-correlation effects of the conduction process. We also consider that the flux through each xylem element is non ideal due to the intricacies related to the internal architecture of the conduction vessels \cite{Tyree2002,Landau1987}. In addition, we assume the Leonardo's seminal hypothesis as valid, that is, the Leonardo's rule is a consequence of the sap flux conservation \cite{Capra2013}. In fact, by attending these requirements we will show that the power law exponent is three and that the sap flux is optimized (ideal fluid behavior) as a consequence of the collective behavior of the xylem elements. 

Our starting point is the mentioned Leonardo's hypothesis \cite{Capra2013}. In specific, if the stem is bifurcated in two daughter branches we can established a relationship between their corresponding fluxes as follows

\begin{equation}
Q=Q_1+Q_2,
\label{eq:1}
\end{equation}

\noindent where $Q$ is the flux of the main branch and $Q_1$ and $Q_2$ are the fluxes of the daughter branches. Now, by assuming that the flow through individual xylem elements is of non ideal nature, that is, it obeys the Hagen-Poseuille law \cite{Tyree2002,Landau1987}, it is possible to establish a relationship between the flux ($q$) and radius ($r$) of the xylem elements as, 

\begin{equation}
q(r) \propto r^4,
\label{eq:2}
\end{equation}

\noindent where the proportionality constant (not shown) comes in terms of intrinsic characteristics of the fluid. For the moment, we consider that the constant is not essential for our derivation. 

\noindent The effective flux through branches can be computed by averaging over all xylem elements

\begin{equation}
Q(R)=\int_0^R f(r) q(r) dr,
\label{eq:3}
\end{equation}

\noindent where $f(r)$ represents the size distribution of xylem elements. By considering a power law distribution

\begin{equation}
f(r)=r^{-x},
\label{eq:4}
\end{equation}

\noindent we can obtain a relationship between the flux and radius of a branch as

\begin{equation}
Q(R) \propto R^{5-x},
\label{eq:5}
\end{equation}

\noindent with $x$ an exponent to be determined. By substituting Eq. (\ref{eq:5}) in Eq. (\ref{eq:1}) we obtain

\begin{equation}
R^{5-x}=R_1^{5-x} + R_2^{5-x},
\label{eq:6}
\end{equation}

\noindent where $R$ is the radius of the main branch, and $R_1$ and $R_2$ are the radii of the daughter branches. 

\noindent Finally, according to the Leonardo's rule the radii of tree branches are related in Pythagorean fashion

\begin{equation}
R^{\Delta}=R_1^{\Delta}+R_2^{\Delta}.
\label{eq:7}
\end{equation} 

\noindent Considering that the Leonardo's exponent is approximately $\Delta \approx 2$, and by comparing Eqs. (\ref{eq:6}) and (\ref{eq:7}) we obtain a power law exponent $x \approx 3$. 

\section{Field work: Measuring steam and branches}
The first task that we carried out it was the corroboration of the Leonardo's rule by directly measuring the stem-branches radii of four different trees species. We measured the stem-branches of the first bifurcation of \textit{Quercus rugosa}, \textit{Eucalyptus camaldulensis}, \textit{Schinus molle} and \textit{Prunus serotina}. We have followed the measuring protocol of Aratsu \citep{aratsu1998}. In particular, we carefully choose healthy trees, with no damage in the branches and with an evident bifurcation. In addition, we have measured the stem and the daughter branches at a reasonable distance from the bifurcation to have reliable data, particularly, to avoid the possible data dispersion associated to the natural branch thickening around the bifurcation. We have considered 91, 50, 48 and 58 individuals for \textit{Quercus rugosa}, \textit{Eucalyptus camaldulensis}, \textit{Schinus molle} and \textit{Prunus serotina}, respectively. 

In order to corroborate the Leonardo's rule we will assume that the perimeter of the stem and daughter branches corresponds to the one of a circle. So, the Pythagorean relationship between radii Eq. (\ref{eq:7}) is also valid for the perimeters

\begin{equation}
\frac{P_1^{\Delta}}{P^{\Delta}} + \frac{P_2^{\Delta}}{P^{\Delta}}=1.
\label{eq:8}
\end{equation}

\noindent Once the perimeters are measured we can compute the Leonardo's exponent $\Delta$. This process is performed for each individual, to finally obtain a net Leonardo's exponent by averaging over all individuals of each tree specie. In the case of \textit{Quercus rugosa}, \textit{Eucalyptus camaldulensis} and \textit{Schinus molle} the perimeters are determined with a measuring tape of one meter length and an experimental error of $\pm 1.0$ mm due to the individuals are adults. For \textit{Prunus serotina} we have used a digital Vernier with a measuring range of 150 mm and a precision of 0.1 mm. In this case the individuals are young trees. For more details about the characteristics of the trees see the supplementary information. 

\section{Experimental work: Maceration process}
The second task that we performed was to unravel the trees xylem elements through the so-called maceration process \cite{fonnegra1978}. This process is recommended for young trees, thus, we implemented it only in \textit{Prunus serotina}. In specific, we macerated three individual according to the following protocol: 

\begin{enumerate}

\item For each individual we prepare a 300 mL maceration solution, of which 120 mL are distilled water, 30 mL are hydrogen peroxide and the remaining 150 mL correspond to glacial acetic acid. 

\item With a handsaw we take pieces of the tree right above and below the bifurcations by cutting the branches transversely. In total we collect five pieces, three at the first bifurcation stage and two more at the second stage. From each piece we extract fine slices with a scalpel. 

\item The fine slices of each piece are placed in a sterile flask with 60 mL of maceration solution. The five flasks of each individual are sealed and introduced in an incubation oven for a five day period at a temperature of 60$^{\circ}$C. 

\item After the time in the incubation oven each flask sample is subjected to three rinses in order to drain the glacial acetic acid. The remaining material is let to rest a whole night in distilled water. 

\item Finally, the distilled water is drained and the remaining solution centrifuge to obtain (separate) the xylem elements. 

\end{enumerate}

\begin{figure}
 \includegraphics[width=\textwidth]{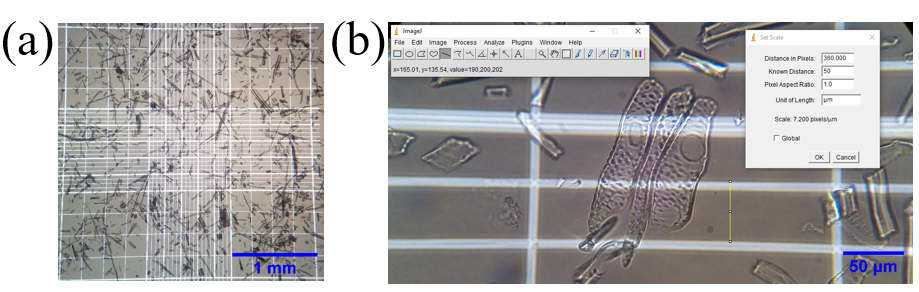}
 \caption{\label{FigPicsCE10X400K} Images of the xylem elements after the maceration process. (a) 10X magnification of a region of the Neubauer chamber, in which we can appreciate the xylem elements dispersed throughout the Neubauer chamber grid. (b)The same as (a), but under a 40X magnification. At the center we can see xylem structures known as xylem vessels \cite{raven2004}.}
 \end{figure}

After the maceration process we proceed to observe the xylem elements. In order to do so, we put the resulting macerated solution in a 0.100 mm depth Neubauer chamber. Specifically, we put a drop of the solution in each of the squares of the chamber, protecting them with a cover glass. Once the sample is ready it is placed in an optical microscope under 100X magnification. We then proceed to scan from left to right and from up to down the observing area by taking pictures of it in the process, see Fig. \ref{FigPicsCE10X400K}. The pictures are processed with the software ImageJ. We consider two methods to obtain the size distribution of the xylem elements. One in which we measure directly the diameter of the xylem elements and other in which the images are black and white contrasted to measure the shaded areas of the xylem elements. More details of these methods can be found in the supplementary material. 

\section{Results}
In this section we present the main results of the field and experimental work. Most importantly, we contrast the resulting size distribution of the xylem elements after the maceration process with the statistical hydraulic model outcomes. 	

Regarding the field work, in table \ref{tbl1} we show the Leonardo's exponent of the tree species studied. The exponents are obtained by computing the zeros of Eq. (\ref{eq:8}) and averaging the exponents of all individuals of each tree specie. The standard error of the exponents is also obtained and shown in table \ref{tbl1}. As we can see the Leonardo's exponent for all tree species lies in the well-known range $1.8 < \Delta < 2.3$ \cite{eloy2011}. It is worth noting that the exponent for adult trees is above 2, while the corresponding one to young trees (\textit{Prunus serotina}) is below 2. As the Leonardo's exponent depends on the tree specie, we also expect that the size distribution of xylem elements does. In fact, in table \ref{tbl1} we also show the expected exponents for the xylem elements size distribution based on our statistical hydraulic model outcomes. As the only tree specie that was subjected to the maceration process was \textit{Prunus serotina} we expect a size distribution exponent $x \approx 3.11$ in the experimental work. 

 \begin{table}%[H] add [H] placement to break table across pages
 \caption{\label{tbl1} Leonardo's exponent of the tree species analyzed in the field work. The expected exponent for the size distribution of the xylem elements is also included for each specie.}
 \begin{ruledtabular}
 \begin{tabular}{cccc}
 Regular name & Scientific name & Leonardo's exponent & Size distribution  \\
 & & & expected exponent \\
 \hline
 Pirul & \textit{Schinus molle} & $2.10 \pm 0.08$ & $2.90 \pm 0.08$ \\
 Eucalyptus & \textit{Eucalyptus camaldulensis} & $2.32 \pm 0.10$ & $2.68 \pm 0.10$  \\
 White Oak & \textit{Quercus rugosa} &  $2.25 \pm 0.06$ & $2.75 \pm 0.06$ \\
 Capulin & \textit{Prunus serotina} & $1.89 \pm 0.07$ & $3.11 \pm 0.07$
 \end{tabular}
 \end{ruledtabular}
 \end{table}
 
\begin{figure}
 \includegraphics[width=\textwidth]{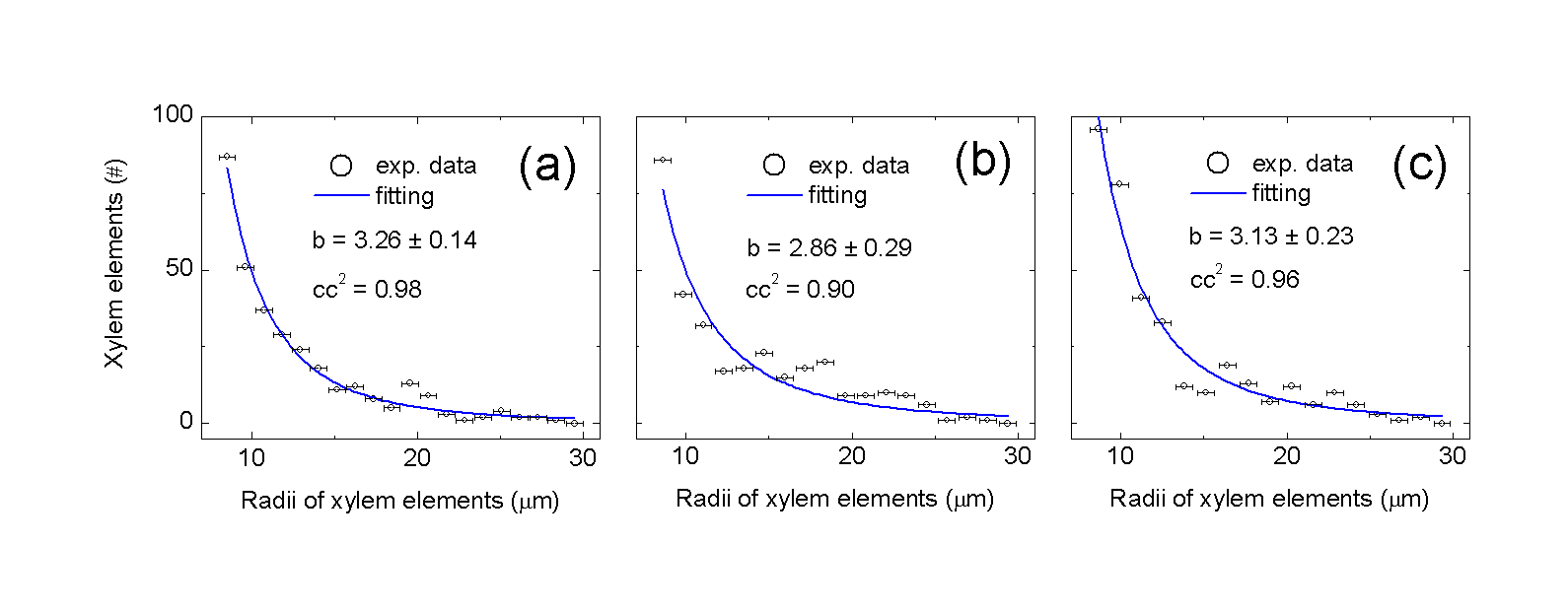}
 \caption{\label{FigEWR1} Size distribution of the xylem elements of the three \textit{Prunus serotina} individuals. The size distributions were determined by directly measuring the diameters of the xylem elements. The solid-blue curves correspond to the power law function fitting. $b$ represents the exponent of the power law function $f(r)=ar^{-b}$. The correlation coefficient (cc$^2$) for each individual is also shown, being best for the first individual. In the three cases about 300 xylem elements were measured.}
 \end{figure}

 Regarding the experimental work, we firstly show the results of measuring directly the diameters of the xylem elements after the maceration process. The size distributions for the three \textit{Prunus serotina} individuals are shown in Fig. \ref{FigEWR1}. As we can see the size of the xylem elements ranges from 8 $\mu$m to 30 $\mu$m. Furthermore, small size xylem elements are the vast majority, resulting in a power law distribution for all individuals. In fact, the best fitting function for the experimental data is of the form $f(r)=ar^{-b}$, see the solid-blue curves in Fig. \ref{FigEWR1}. In all cases the correlation coefficient is over 90 \%, indicating a fairly good fitting for the power law function. We also determine the relative error of the resulting exponent with respect to the expected one. The resulting exponents for the three individuals as well as its corresponding relative errors are shown in table \ref{tbl2}. As we can notice the exponent of the third individual ($b \approx 3.13$) is closest to the expected one ($x \approx 3.11$), with a relative error of 0.64 \%. Here, it important to mention that measuring directly the diameter of the xylem elements is time demanding and consequently impractical to process all images of the three  \textit{Prunus serotina} individuals. So, the exponents that we obtained are the result of processing a reduced number of xylem elements images. 
 
\begin{table}%[H] add [H] placement to break table across pages
 \caption{\label{tbl2}Size distribution exponent for the \textit{Prunus serotina} individuals by measuring directly the diameters of the xylem elements. The exponent relative error is also presented. The exponent is calculated by fitting the experimental data to the power law function $f(r)=ar^{-b}$. The relative error is computed by comparing the exponent obtained in the experimental work ($b$) with the expected one ($x$) according to the field work and the statistical hydraulic model: $\epsilon=\frac{b-x}{x} \times 100$.}
 \begin{ruledtabular}
 \begin{tabular}{ccc}
 Capulin individuals & Measured exponent ($b$) & Relative error ($\epsilon$)  \\
 \hline
First  & 3.26 & 4.80 \% \\
Second  & 2.88 & 7.39 \% \\
Third  & 3.13 & 0.64 \% 
 \end{tabular}
 \end{ruledtabular}
 \end{table}
 
In order to overcome this obstacle, we have used an alternative method based on black and white contrast the xylem elements images. This method allows us to measure in an automatic way the shaded areas of the xylem elements and to determine indirectly the corresponding diameters. More details can be found in the supplementary information. In Fig. \ref{FigEWR2} we show the outcomes of this alternative method for the first  \textit{Prunus serotina} individual. In this case we have measured the area of about $10,000$ xylem elements. As we can notice the size of xylem elements ranges from 20 $\mu$m to 180 $\mu$m, being the vast majority small size xylem elements. The experimental data fit quite well to a power law function (solid-blue curve), with a size distribution exponent $b \approx3.08$ and a 99 \% correlation coefficient. In addition, the relative error of the size distribution exponent is 0.96 \%. 

%The results for the other Prunus Salicifolia individuals are similar, however the distribution function exponent and the relative error are not as good as for the first individual. The corresponding results %can be found in the supplementary material. 

\begin{figure}
 \includegraphics[width=\textwidth]{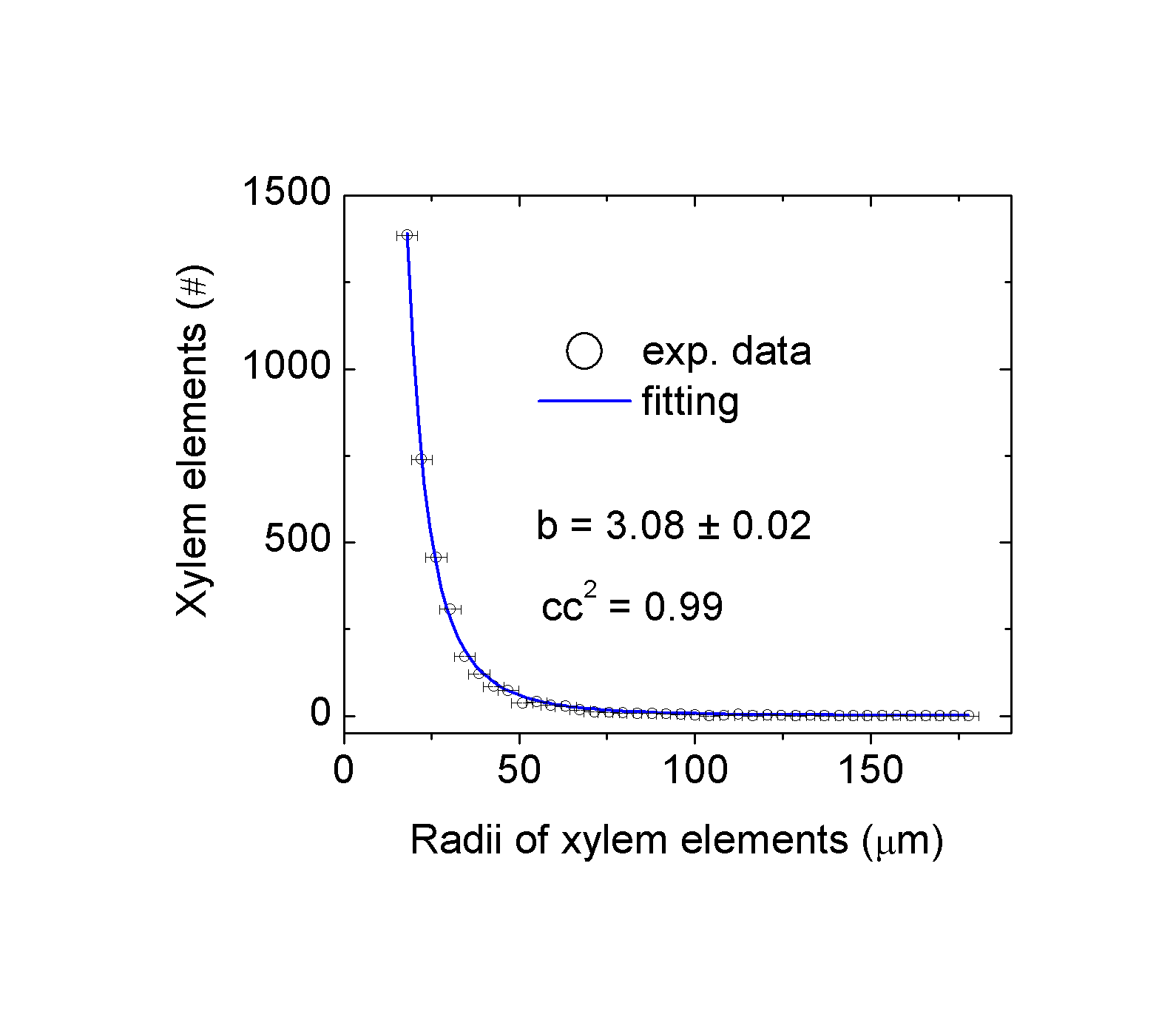}
 \caption{\label{FigEWR2} Size distribution of the xylem elements of the first  \textit{Prunus serotina} individual by measuring the areas of the xylem elements according to the black-white contrast method. The solid-blue curve correspond to the power law function fitting. In this case about 10,000 xylem elements were measured.}
 \end{figure}

\section{Discussion}
The first aspect that we want to discuss is related to the characteristics of the xylem elements. It is well known that as trees grow some xylem elements lose its conductive ability, as they fill with resin, giving place to the so-called heartwood \cite{Tyree2002}. At the same time new xylem elements arise, keeping the tree conducting capability. The elements that compose the xylem have in its walls lignin \cite{Tyree2002,donaldson2001}, a substance that makes them rigid and woody, providing mechanical stability and support to the tree. Both conducting and non-conducting elements contribute to the thickness of the stem and branches. With the maceration process what we obtain is a disperse sample with conducting and non-conducting elements. Our first image processing method allows us to choose xylem structures with conducting capabilities such as tracheids and vessel elements. With the black and white contrast method we measure the areas of conducting and non-conducting elements. Though, heartwood structures are not participating in the conduction process, they are relevant in the determination of the size distribution because they contribute to the thickness of the stem and branches. Under this context, we expect results in better agreement with respect to the predicted ones with the black and white contrast method. In fact, this is what happened with the first individual. After considering conducting and non-conducting elements the exponent of the size distribution ($b \approx 3.08$) results in close agreement with respect to the predicted one ($x \approx 3.11$) with a relative error of 0.96 \%.  

The second aspect that we want to address it is about the type of distribution function that was obtained. In specific, a power law distribution function alike to the so-called  L\'evy-type distribution functions \cite{sato1999,barndorff2001,feller1968}. In fact, these distribution functions are characterized by memory and correlation effects. Taking into account that xylem structures form a complex conducting network \cite{Tyree2002}, in which memory and correlation effect are plausible, we consider that it is totally reasonable to obtain a power law distribution for the xylem elements. In fact, a power law distribution, in which small size xylem elements dominate, optimized the conduction process by changing the non-ideal fluid behavior of individual xylem elements to an ideal one after summing over all xylem elements. Actually, at first sight, it is counterintuitive that a network of non-idea fluid xylem elements can optimized the conduction process. However, what really matters it is the collective effect of all conducting structures rather than its individual behavior. This intricate organization of the xylem elements it is likely to be related to the space-filling maximization as in the case of other biological complex networks \cite{viswanathan2008,gillooly2001,humphries2012}. Our results are also indicative that the generation of the xylem elements, from the meristematic tissue called Cambium \cite{larson1994}, is taking place in such a way that small radius xylem structures dominate over large ones, giving rise to a power law distribution. 

Finally, we want to discuss the results obtained with the black and white image-processing method. In principle, it is possible that with the maceration process a lot of conducting structures be broken. If so, what we are obtaining with the black and white image-processing method would be the distribution function of a rupture process \cite{sotolongo1994,sotolongo1996}, as we are measuring automatically all structures in our sample. However, the exponent of the distribution function associated to a rupture process it is well known \cite{sotolongo1994,sotolongo1996} and it is not the one that we obtained experimentally. Furthermore, taking into account the similarities of the distribution functions obtained with both image-processing methods we truly believe that the results obtained with the black and white contrast method correspond to the conducting elements rather than a rupture process. 

\section{Conclusions}
In summary, we have assessed the Leonardo's rule of tree branching from the hydraulic standpoint. We have proposed a model based on the flux conservation, the Hagen-Poseuille law and a xylem elements size-distribution of the L\'evy-type. The model tells us that when the Leonardo's rule is fulfilled the size-distribution presents a power law exponent of three. Furthermore, the total flux is optimized, going from a non-ideal fluid for each xylem element to an idea fluid after summing over all xylem elements. We consider that this optimization process is reasonable because the minerals transported by xylem have to reach all parts of the three to make the photosynthesis process possible. We have corroborated the Leonardo's rule by directly measuring the stem-branches diameter of four different tree species: \textit{Quercus rugosa}, \textit{Eucalyptus camaldulensis}, \textit{Schinus molle} and \textit{Prunus serotina}. For the latter we carried out a maceration process in order to unveil the size and number of xylem elements. We found a power law distribution for the xylem elements with an exponent close to three. This value is in close agreement with our theoretical prediction, implying that the seminal hypothesis of Leonardo, the flux conservation is directly related in the trees branching, is valid.

\begin{acknowledgments}
P.V.-M. acknowledges to CONACYT-Mexico for the scholarship for graduate studies. 
\end{acknowledgments}

\bibliography{BiblioLeonardo.bib}

\end{document}